\documentclass[%
 reprint,
 amsmath,amssymb,
 aps,
]{revtex4-2}

\usepackage{graphicx}
\usepackage{dcolumn}
\usepackage{color}
\usepackage{bm}

\begin{document}

\preprint{APS/123-QED}

\title{Scattering theory for stationary materials with $\mathcal{PT}$ symmetry}

\author{P. A. Brand\~ao}
\email{paulo.brandao@fis.ufal.br}
 \affiliation{Instituto de F\'isica, Universidade Federal de Alagoas, Macei\'o, 57072-900, Brazil.}
\author{O. Korotkova}%
 \email{korotkova@physics.miami.edu}
\affiliation{Department of Physics, University of Miami, 1320 Campo Sano Dr, Coral Gables, FL 33146, USA.}%

\date{\today}

\begin{abstract}
A theoretical framework is developed for scattering of scalar radiation from stationary, three-dimensional media with correlation functions of scattering potentials obeying $\mathcal{PT}$-symmetry. It is illustrated that unlike in scattering from deterministic $\mathcal{PT}$ symmetric media, its stationary generalization involves two mechanisms leading to symmetry breaking in the statistics of scattered radiation, one stemming from the complex-valued medium realizations and the other - from the complex-valued degree of medium's correlation.
\end{abstract}

\maketitle


\section{Introduction}

The discovery made by Bender and Boettcher \cite{1} on the possibility of physical systems with complex-valued Hamiltonians to possess real-valued spectra, under conditions of the $\mathcal{PT}$ symmetry, set the basis for the explosive growth of non-Hermitian quantum mechanics \cite{2}. This generalization from Hermitian to non-Hermitian quantum mechanics acquired a solid foundation after the introduction of a new definition for the inner product of the Hilbert space \cite{3}. It was soon realized that $\mathcal{PT}$ symmetry is actually a special case of a more general class of pseudo-Hermitian Hamiltonians that yield real eigenvalues \cite{4}. Electromagnetics and optics, in particular, has soon benefited from such a generalization in view of the analogy between the paraxial wave equation and the time-dependent Schr\"odinger equation, with manifestation in the areas of wave-guiding \cite{5}, unidirectional crystal invisibility \cite{6}, balanced lasing and anti-lasing \cite{7}, structured light emission \cite{8}, beam dynamics in periodic lattices \cite{9}, among others.   

In the realm of physical optics, the $\mathcal{PT}$ symmetry of a deterministic material translates to Hermiticity of its complex-valued index of refraction or, equivalently, scattering potential \cite{10}. For static media, only spatial Hermiticity at a fixed frequency must be imposed. Physically  this implies the perfect balance between gain and loss centers constituting the medium. Hence the symmetry in the complex-valued index of refraction must be considered with precaution, for it can be geometrically symmetric but not Hermitian, implying imbalance between the gain and loss contributions. In fact, the geometrically symmetric (unbalanced) and the  $\mathcal{PT}$-symmetric (balanced) media are mutually exclusive, with the only exception possible when the imaginary parts of their refractive index, accounting for gain and loss, vanish. 

In this work, we pursue a set of objectives, the first of which is to draw a clear distinction between a static, deterministic, geometrically symmetric medium, which we will term \textit{classic}, and a $\mathcal{PT}$-symmetric medium. The second, and the main aim, is to introduce a class of stochastic, stationary media obtained on taking the spatial correlation over the ensemble of  realizations of the medium obeying $\mathcal{PT}$ symmetry. The detailed analysis of the properties of the new type of  correlation functions, their mathematical modeling and their comparison with those of classic media is then attempted. Our final goal is to develop a theoretical treatment of light scattering from the new type of stationary media, and compare it with that from classic stationary media \cite{11}. The central quantity in the scattering theory involving generally random illumination and random medium is the pair-scattering matrix \cite{12} which characterizes the change by the medium in the correlation along two incident and two scattered directions and which coincides, within the accuracy of the first Born approximation, with the Fourier transform of the scattering potential correlation function. We derive this quantity for the general stationary $\mathcal{PT}$-symmetric media and reveal its remarkable properties. 

In addition, a mathematical model for a Schell-like correlation function of the  $\mathcal{PT}$-symmetric medium with linear phases is developed with the help of the Bochner theorem previously used in optics in connection to field correlations \cite{13}, \cite{14} and classic medium correlations \cite{15} for numerical illustration of the new theory.

Various aspects of stochastic light scattering from deterministic $\mathcal{PT}$-symmetric media were recently discussed in Refs. \cite{16}, \cite{17} and \cite{18}. The results of these studies can be deduced from the general theory developed here in the completely correlated medium limit. To our knowledge the only example of a $\mathcal{PT}$-symmetric stationary medium had previously been considered in \cite{19}, however without development of the general theory.

The paper is organized as follows: the stationary $\mathcal{PT}$-symmetric media and their properties are analyzed in Sec. 2; the general approach for modeling of the novel media is developed in Sec. 3; Sec. 4 presents the general theory for characterization of weak scattering from such media; the analytical examples relating to scattering of Schell-like $\mathcal{PT}$-symmetric media are given in Sec. 5 and the concluding remarks are provided in Sec. 6.

\section{Complex material correlation functions with classic and $\mathcal{PT}$ symmetries}

\begin{figure}[ht]
    \centering
    \includegraphics[width=0.45\textwidth]{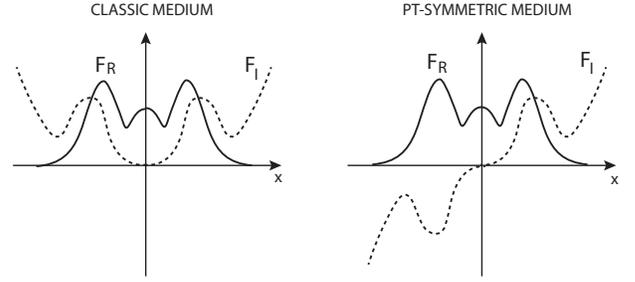}
    \caption{Real and imaginary parts of typical classic (left) and $\mathcal{PT}$-symmetric (right) media's scattering potentials, along a single Cartesian axis, say $x$.}
    \label{fig0}
\end{figure}

We will confine our attention only to a (wide-sense) statistically stationary medium distributed in a three-dimensional region of space and being symmetric, in a certain sense, with respect to its  geometrical center, say $\vec{r}=0$. Let us first consider the case of an unbalanced stationary medium in which the realizations of the complex-valued index of refraction 
\begin{equation}
n(\vec{r},\omega)=n_r(\vec{r},\omega)+in_i(\vec{r},\omega)
\end{equation}    
at position $\vec{r}$ and angular frequency $\omega$, obey geometrical symmetry relations
\begin{equation}
n_r(-\vec{r},\omega)=n_r(\vec{r},\omega), \hspace{1cm} n_i(-\vec{r},\omega)=n_i(\vec{r},\omega). 
\end{equation}   
We will refer to such media as having classic symmetry or just \textit{classic} and use them throughout the paper to set up the contrast with $\mathcal{PT}$-symmetric media that we will introduce below. The realizations of the scattering potential defined as \cite{11} 
\begin{equation}
F(\vec{r},\omega) = (k^2/4\pi^2)[n^2(\vec{r},\omega) - 1],
\end{equation}
where $k$ is the radiation's wave number, then must be symmetric about the center as well: 
\begin{equation}\label{FCL}
F_{CL}(-\vec{r},\omega)=F_{CL}(\vec{r},\omega),
\end{equation}
where subscript $CL$ is used to denote classic (unbalanced) medium. See Fig. \ref{fig0} (left) for visualization of real and imaginary parts of $F_{CL}$.

This immediately implies that the spatial correlation function of the scattering potential in Eq. \eqref{FCL} \cite{11}
\begin{equation}\label{C}
C_{CL}(\vec{r}_1,\vec{r}_2,\omega)=\langle F^*_{CL}(\vec{r}_1,\omega)F_{CL}(\vec{r}_2,\omega)\rangle_m,
\end{equation}
of the classic medium, where subscript $m$ denotes averaging over medium realizations, must meet condition 
\begin{equation}\label{Cclass}
C_{CL}(-\vec{r}_1,-\vec{r}_2,\omega) = C_{CL}(\vec{r}_1,\vec{r}_2,\omega),
\end{equation}
i.e., must also be symmetric with respect to the medium's center $\vec{r}=0$. We will also refer to such stationary media as \textit{classic}. As we will demonstrate, due to their inherent geometrical symmetry, such media constitute a solid comparison tool with the $\mathcal{PT}$-symmetric media. Indeed, random media with spatially asymmetric, real-valued correlation functions, can also be shown to lead to broken symmetries in the statistics of scattered radiation \cite{20}. 

On the other hand, a $\mathcal{PT}$-symmetric material is described by a complex refractive index whose real and imaginary parts must satisfy relations
\begin{equation}
    n_r(-\vec{r},\omega) = n_r(\vec{r},\omega), \quad n_i(-\vec{r},\omega) = -n_i(\vec{r},\omega).
\end{equation}
Then the scattering potential of such a medium is also $\mathcal{PT}$-symmetric and satisfies condition
\begin{equation}\label{PTF}
    F^*_{PT}(-\vec{r},\omega) = F_{PT}(\vec{r},\omega).
\end{equation}
In view of Eq. \eqref{PTF} the two-point correlation function of the $\mathcal{PT}$-symmetric scatterer 
\begin{equation}\label{C}
C_{PT}(\vec{r}_1,\vec{r}_2,\omega)=\langle F^*_{PT}(\vec{r}_1,\omega)F_{PT}(\vec{r}_2,\omega)\rangle_m,
\end{equation}
must satisfy relation
\begin{equation}\label{CPT}
    C^*_{PT}(-\vec{r}_1,-\vec{r}_2,\omega) = C_{PT}(\vec{r}_1,\vec{r}_2,\omega).
\end{equation}
We will term the media with correlation in Eq. \eqref{CPT} $\mathcal{PT}$-\textit{symmetric stationary media}. 

In order to avoid possible misconceptions we will now  outline some other similarities and differences between the two media types, based on comparison between Eqs. (\ref{Cclass}) and (\ref{CPT}). First, switching the order of spatial arguments in any complex-valued correlation function yields $C(\vec{r}_2,\vec{r}_1,\omega)=C^*(\vec{r}_1,\vec{r}_2,\omega)$ which implies with the use of Eqs. \eqref{Cclass} and  \eqref{CPT} that  
\begin{equation}\label{switch}
\begin{split}
C_{CL}(\vec{r}_2,\vec{r}_1,\omega) &= C_{CL}^*(-\vec{r}_1,-\vec{r}_2,\omega), \\
C_{PT}(\vec{r}_2,\vec{r}_1,\omega) &= C_{PT}(-\vec{r}_1,-\vec{r}_2,\omega).
\end{split}
\end{equation}
Second, on defining the measure of the \textit{strength of scattering potential} for the two media by the expressions
\begin{equation}
I_{CL}(\vec{r},\omega) = C_{CL}(\vec{r},\vec{r},\omega), \quad I_{PT}(\vec{r},\omega) = C_{PT}(\vec{r},\vec{r},\omega),
\end{equation}
we find that they are given by the same formula:
\begin{equation}\label{ICL}
I_{\alpha}(\vec{r},\omega) =
 \langle|F(\vec{r},\omega) |^2\rangle_m \quad (\alpha = CL,PT),
\end{equation}
being, due to Eq. (\ref{Cclass}) and (\ref{CPT}), the  real-valued and even functions 
\begin{equation}
I_{\alpha}(-\vec{r},\omega)=I_{\alpha}(\vec{r},\omega).
\end{equation}

In order to highlight the distinctive feature of stationary $\mathcal{PT}$-symmetric media we introduce a new quantity, \textit{anti-strength of scattering potential}, which is an analog of a cross-correlation function used sometimes in connection to optical fields (c.f. Ref. \cite{21})
\begin{equation}\label{NrwPT}
\begin{split}
N_{PT}(\vec{r},\omega)=& C_{PT}(-\vec{r},\vec{r},\omega)=\langle F^2(\vec{r},\omega) \rangle_m,
\end{split}
\end{equation}
and note that $\text{Im}[N_{PT}(\vec{r},\omega)] = (k^4/4\pi^4)\langle n_rn_i(n_r^2 - n_i^2 - 1) \rangle$ relates to the amount of asymmetry that a $\mathcal{PT}$-symmetric medium introduces on scattering, while $\text{Im}[N_{PT}(0,\omega)]=0$. For classic media this quantity can also be defined but coincides with $I_{CL}$.

More generally, the \textit{degrees of potential correlation} defined at any points $\vec{r}_1$ and $\vec{r}_2$ by expressions 
\begin{equation}\label{mualpha}
\mu_{\alpha}(\vec{r}_1,\vec{r}_2,\omega)=\frac{C_{\alpha}(\vec{r}_1,\vec{r}_2,\omega)}{\sqrt{I(\vec{r}_1,\omega)}{\sqrt{I(\vec{r}_2,\omega)}}},  \quad  (\alpha=CL,PT)
\end{equation}
 are seen to satisfy the same relations as $C_{CL}$ and $C_{PT}$, respectively, but their values are bound to the unit circle of the complex plane. Equation \eqref{mualpha} has a particularly simple form at symmetric points:
\begin{equation}
\mu_{\alpha}(-\vec{r},\vec{r},\omega)=\frac{N_{\alpha}(\vec{r},\omega)}{I(\vec{r},\omega)}, \quad  (\alpha=CL,PT).
\end{equation}
This immediately implies that while a classic medium is fully correlated (for such points), viz., 
\begin{equation}
\mu_{CL}(-\vec{r},\vec{r},\omega)=1,
\end{equation}
the $\mathcal{PT}$-symmetric medium can have any, generally complex-valued, correlation state:
\begin{equation}
\mu_{PT}(-\vec{r},\vec{r},\omega)=\frac{\langle F^2(\vec{r},\omega) \rangle_m}{\langle |F(\vec{r},\omega)|^2 \rangle_m}.
\end{equation}

\section{Modeling of $\mathcal{PT}$-symmetric media correlation functions}

The Bochner theorem of functional analysis was employed in Ref. \cite{15} for modeling of novel three-dimensional scattering media (see also \cite{13}, \cite{14} for its application to stationary radiation). We will now explore the application of this idea  specifically to the $\mathcal{PT}$-symmetric stationary media. For $C_{PT}(\vec{r}_1,\vec{r}_2,\omega)$ to represent a genuine correlation function, it is sufficient to write it as  
\begin{equation}\label{Bochner}
    C_{PT}(\vec{r}_1,\vec{r}_2,\omega) = \int p(\vec{v},\omega)H^*_{PT}(\vec{r}_1,\vec{v},\omega)H_{PT}(\vec{r}_2,\vec{v},\omega)d^3v,
\end{equation}
where $p(\vec{v},\omega) \geq 0$,  $H_{PT}(\vec{r},\vec{v},\omega)$ is a complex-valued function and the integration extends over the three-dimensional space of vector $\vec{v}$. Equation \eqref{CPT} imposes some restrictions on the possible forms that $H_{PT}(\vec{r},\vec{v},\omega)$ can assume:
\begin{multline}
    \int p(\vec{v},\omega)H_{PT}(-\vec{r}_1,\vec{v},\omega)H^*_{PT}(-\vec{r}_2,\vec{v},\omega)d^3v \\ 
    = \int p(\vec{v},\omega)H^*_{PT}(\vec{r}_1,\vec{v},\omega)H_{PT}(\vec{r}_2,\vec{v},\omega)d^3v.
\end{multline}
For this relation to be satisfied it is \textit{sufficient} that
\begin{multline}
    H_{PT}(-\vec{r}_1,\vec{v},\omega)H^*_{PT}(-\vec{r}_2,\vec{v},\omega) \\
    = H^*_{PT}(\vec{r}_1,\vec{v},\omega)H_{PT}(\vec{r}_2,\vec{v},\omega), 
\end{multline}
or
\begin{equation}\label{HPT}
    H^*_{PT}(-\vec{r},\vec{v},\omega) = H_{PT}(\vec{r},\vec{v},\omega),
\end{equation}
implying that $H_{PT}(\vec{r},\vec{v},\omega)$ is $\mathcal{PT}$ -symmetric. 
 It is interesting to compare Eq. \eqref{HPT} with that for classic media. 
 On writing 
 \begin{equation}\label{BochnerCL}
    C_{CL}(\vec{r}_1,\vec{r}_2,\omega) = \int p(\vec{v},\omega)H^*_{CL}(\vec{r}_1,\vec{v},\omega)H_{CL}(\vec{r}_2,\vec{v},\omega)d^3v,
\end{equation}
 we find that in view of the first of Eqs. \eqref{switch}, no additional condition is imposed on $H_{CL}$: indeed, it can be any complex-valued function. We stress that condition \eqref{HPT} is only sufficient, in general $H_{PT}$ may have a more general form. 
 
For Schell-like scatterers the correlation class takes the form of the three-dimensional Fourier transform kernel
\begin{equation}
    H(\vec{r},\vec{v},\omega) = a(\vec{r},\omega)\exp(-2\pi i\vec{r}\cdot\vec{v}),
\end{equation}
where $a(\vec{r},\omega)$ must satisfy $a^*(-\vec{r},\omega) = a(\vec{r},\omega)$ provided Eq. \eqref{HPT} holds. Then, Eq. \eqref{Bochner} simplifies as
\begin{equation}\label{CPTFTkernel}
\begin{split}
    C_{PT}(\vec{r}_1,\vec{r}_2&,\omega) = a^*(\vec{r}_1,\omega)a(\vec{r}_2,\omega) \\
    &\times\int p(\vec{v},\omega)\exp[-2\pi i \vec{v}\cdot(\vec{r}_2-\vec{r}_1)]d^3v \\
    &= a^*(\vec{r}_1,\omega)a(\vec{r}_2,\omega)g(\vec{r}_d,\omega)\otimes g(\vec{r}_d,\omega) \\
     &= \sqrt{I_{PT}(\vec{r}_1,\omega)} \sqrt{I_{PT}(\vec{r}_2,\omega)}\\
    &\times \exp[-i\psi(\vec{r}_1,\omega)]\exp[i\psi(\vec{r}_2,\omega)] \\
    &\times\mu_{PT}(\vec{r}_d,\omega) ,
\end{split}
\end{equation}
where $\sqrt{I_F(\vec{r},\omega)} = |a(\vec{r},\omega)|$, $\psi(\vec{r},\omega)=\arg[a(\vec{r},\omega)]$, $\otimes$ denotes the 3D convolution, $\vec{r}_d = \vec{r}_2-\vec{r}_1$, $\mu_{PT}(\vec{r}_d,\omega) = g(\vec{r}_d,\omega)\otimes g(\vec{r}_d,\omega)$ and $g(\vec{r}_d,\omega)$ is given by integral
\begin{equation}
    g(\vec{r}_d,\omega) = \int \sqrt{p(\vec{v})}\exp(-2\pi i \vec{v}\cdot \vec{r}_d) d^3v.
\end{equation}

There are two fundamental ways in which $C_{PT}$ in Eq. \eqref{CPTFTkernel} can possess nontrivial phase: function $\psi$ can be non-trivial or $\mu_{PT}$ can be complex-valued. This results in a qualitatively different far-field spectral density distributions. In the former case an originally deterministic $\mathcal{PT}$-symmetric medium is randomized, resulting in partial ``blurring'' of the scattered spectral density, as compared with that for $\mu_{PT}=1$. In the latter case the medium is not $\mathcal{PT}$-symmetric and if $\mu_{PT}=1$ it would scatter to spectral density symmetric about the $z$-axis. However, if $|\mu_{PT}| \rightarrow 0$ and complex-valued it produces $\mathcal{PT}$-symmetry like effects in which Fourier transform of $|\mu_{PT}|$ determines the scattered spectral density profile and $\arg(\mu_{PT})$ determines the off-axis shift. Both conditions can hold simultaneously, leading to a much more complex scattering outcome. In particular, two mechanisms can annihilate the asymmetries produced by them individually.

\section{Scattering theory for stationary $\mathcal{PT}$-symmetric media}

Within the validity of the first Born approximation, the cross-spectral density $W^s(r\hat{s}_1,r\hat{s}_2,\omega)$ of radiation scattered to the far zone of a stationary medium is given by integral (\cite{11}, p. 120)
\begin{equation}\label{prop1}
\begin{split}
    W^s(r\hat{s}_1,&r\hat{s}_2,\omega) = \frac{1}{r^2}\int_{V}\int_{V} W^i(\vec{r}_1,\vec{r}_2,\omega)C(\vec{r}_1,\vec{r}_2,\omega) \\
    &\times\exp[-i k(\hat{s}_2\cdot\vec{r}_2 - \hat{s}_1\cdot\vec{r}_1)]d^3r_1 d^3r_2,
\end{split}
\end{equation}
where $W^i(\vec{r}_1,\vec{r}_2,\omega)$ is the cross-spectral density of the incident field. We can express Eq. \eqref{prop1} as
\begin{equation}\label{prop2}
\begin{split}
    W^s(r\hat{s}_1,&r\hat{s}_2,\omega) \\&= \frac{1}{r^2} \widetilde{W}^i(\vec{r}_1,\vec{r}_2,\omega)\circledast \widetilde{C}(\vec{r}_1,\vec{r}_2,\omega)\Biggr|_{(-k\hat{s}_1,
    k\hat{s}_2,\omega)},
    \end{split}
\end{equation}
where tilde stands for three-dimensional Fourier transform and $\circledast$ denotes convolution in six dimensions. This expression indicates that $\widetilde{C}$ plays the crucial part in characterizing the redistribution of energy  along various incident and scattered directions. See also \cite{12} for more general expressions for $W^s$ involving $\widetilde{C}$ for the intermediate scattered field, relating to the pair-scattering matrix. Expression \eqref{prop2} substantially simplifies under the assumption  that the incident field is a polychromatic plane wave propagating along direction $\hat{s}_0$, and, hence, having the cross-spectral density of the form 
\begin{equation}\label{PW}
    W^i(\vec{r}_1,\vec{r}_2,\omega) = S^i(\omega)\exp[i k\hat{s}_0\cdot(\vec{r}_2 - \vec{r}_1)],
\end{equation}
 where $S^i(\omega)$ is the (position-independent) spectral density. After substituting from Eq. \eqref{PW} into \eqref{prop2} we obtain 
\begin{equation}\label{Ws}
\begin{split}
    W^s(r\hat{s}_1,r\hat{s}_2,\omega) =\frac{S^i(\omega)}{r^2} \widetilde{C}(-\vec{K}_1,\vec{K}_2,\omega),
    \end{split}
\end{equation}
where  
\begin{equation}
    \vec{K}_1 = k(\hat{s}_1 - \hat{s}_0) \hspace{0.5cm} \text{and} \hspace{0.5cm} \vec{K}_2 = k(\hat{s}_2 - \hat{s}_0)
\end{equation}
are the momentum transfer vectors characterizing  scattering from incident direction $\hat{s}_0$ to outgoing direction $\hat{s}_j$, $j=1,2$. We note that in cases when $W^i$ involves more than one direction, the momentum transfer vectors have more general form: $\vec{K}_1 = k(\hat{s}_1 - {\hat{s}'}_1)$ and $\vec{K}_2 = k(\hat{s}_2 - {\hat{s}'}_2)$, i.e., they depend on two incident and two scattered directions.

Let us now analyze $\tilde{C}(-\vec{K}_1,\vec{K}_2,\omega)$ in detail:
\begin{equation}\label{tildeC}
    \begin{split}
        \tilde{C}(-\vec{K}_1,\vec{K}_2,\omega)& =\langle \tilde{F}_r(-\vec{K}_1,\omega)\tilde{F}_r(\vec{K}_2,\omega) \rangle_m \\
        &+ \langle\tilde{F}_i(-\vec{K}_1,\omega)\tilde{F}_i(\vec{K}_2,\omega) \rangle_m \\
        &+i\big[ \langle \tilde{F}_r(-\vec{K}_1,\omega)\tilde{F}_i(\vec{K}_2,\omega) \rangle_m \\
        &- \langle \tilde{F}_i(-\vec{K}_1,\omega)\tilde{F}_r(\vec{K}_2,\omega) \rangle_m \big].
    \end{split}
\end{equation}
For classic medium both $F_r(\vec{r},\omega)$  and $F_i(\vec{r},\omega)$ are real-valued and even, hence both 
$\tilde{F}_r(\vec{K},\omega)$ and $\tilde{F}_i(\vec{K},\omega)$ are real valued and even, 
$\tilde{F}_r(-\vec{K},\omega) = \tilde{F}_r(\vec{K},\omega)$ [$\tilde{F}_i(-\vec{K},\omega) = \tilde{F}_i(\vec{K},\omega)$]. This implies that 
\begin{equation}\label{symmetryCL}
\begin{split}
\tilde{C}_{CL}(-\vec{K}_2,\vec{K}_1,\omega)& =\tilde{C}_{CL}^*(-\vec{K}_1,\vec{K}_2,\omega).
\end{split}
\end{equation}
Since all Fourier transforms entering $\tilde{C}_{CL}$ are real functions, it must always be complex-valued if $\vec{K}_1 \neq {\vec{K}_2}$, unless $F_i(\vec{r},\omega)=0$.
For $\mathcal{PT}$-symmetric medium $F_r(\vec{r},\omega)$ [$F_i(\vec{r},\omega)$] is even [odd], hence we have $\tilde{F}_r(-\vec{K},\omega) = \tilde{F}_r(\vec{K},\omega)$ [$\tilde{F}_i(-\vec{K},\omega) = -\tilde{F}_i(\vec{K},\omega)$], and $\tilde{F}_r(\vec{K},\omega)$ is real-valued, while $\tilde{F}_i(\vec{K},\omega)$ is purely imaginary. This implies
\begin{equation}\label{tildeCPT}
\begin{split}
\tilde{C}_{PT}(-\vec{K}_2,\vec{K}_1,\omega)& =\tilde{C}_{PT}(-\vec{K}_1,\vec{K}_2,\omega).
\end{split}
\end{equation}
Further, since $\tilde{F}_i(\vec{K},\omega)$ is a purely imaginary function, Eq. \eqref{tildeCPT} implies that $\tilde{C}_{PT}$ must always be real-valued, even for $\vec{K}_1\neq \vec{K}_2$ and $F_i(\vec{r},\omega)\neq 0$.

Along the same scattered direction $\hat{s}=\hat{s}_1=\hat{s}_2$, $\vec{K}=\vec{K}_1=\vec{K}_2$, and, hence, the spectral density becomes
\begin{equation}
\begin{split}
S^{s}(r\hat{s},\omega)&=\frac{S^i(\omega)}{r^2} \tilde{C}(-\vec{K},\vec{K},\omega) \\& = \frac{S^i(\omega)}{r^2}\tilde{N}(\vec{K},\omega),
\end{split}
\end{equation}
where
\begin{equation}\label{tildeN}
\begin{split}
\tilde{N}&(\vec{K},\omega)=\tilde{C}(-\vec{K},\vec{K},\omega) \\
&=\langle \tilde{F}_r(-\vec{K},\omega)\tilde{F}_r(\vec{K},\omega)\rangle_m +
\langle \tilde{F}_i(-\vec{K},\omega)\tilde{F}_i(\vec{K},\omega)\rangle_m\\&+i[\langle \tilde{F}_r(-\vec{K},\omega)\tilde{F}_i(\vec{K},\omega)\rangle_m-
\langle \tilde{F}_i(-\vec{K},\omega)\tilde{F}_r(\vec{K},\omega)\rangle_m].
\end{split}
\end{equation}
In particular,
\begin{equation}
\begin{split}
        \tilde{N}_{CL}(\vec{K},\omega) &=  \langle \tilde{F}_r^2(\vec{K},\omega) \rangle_m + \langle \tilde{F}_i^2(\vec{K},\omega) \rangle_m\\&
        =\langle| \tilde{F}(\vec{K},\omega) |^2\rangle_m,
        \end{split}
\end{equation}
being the Fourier transform of $I_{CL}(\vec{r},\omega)$ in Eq. \eqref{ICL} and
\begin{equation}
    \begin{split}
        \tilde{N}_{PT}(\vec{K},\omega) &=\langle [\tilde{F}_r(\vec{K},\omega)+i\tilde{F}_i(\vec{K},\omega)]^2 \rangle_m \\&
        =\langle \tilde{F}^2(\vec{K},\omega) \rangle_m,
    \end{split}
\end{equation}
being the Fourier transform of $N_{PT}(\vec{r},\omega)$ in Eq. \eqref{NrwPT}. Both $\tilde{N}_{CL}(\vec{K},\omega)$ and $\tilde{N}_{PT}(\vec{K},\omega)$ are obviously real-valued. By definition, the spectral degree of coherence of the scattered field is
\begin{equation}
    \mu^s(r\hat{s}_1,r\hat{s}_2,\omega) = \frac{W^s(-\vec{K}_1,\vec{K}_2,\omega)}{\sqrt{W^s(-\vec{K}_1,\vec{K}_1,\omega)}\sqrt{W^s(-\vec{K}_2,\vec{K}_2,\omega)}}
\end{equation}
and in view of Eq. \eqref{Ws} it generally yields 
\begin{equation}\label{mus}
    \mu^s(r\hat{s}_1,r\hat{s}_2,\omega) = \frac{\tilde{C}(-\vec{K}_1,\vec{K}_2,\omega)}
    {\sqrt{\tilde{N}(\vec{K}_1,\omega)}\sqrt{\tilde{N}(\vec{K}_2,\omega)}}.
\end{equation}

To summarize the results of this section: (I) for generic vectors $\vec{K}_1$ and $\vec{K}_2$,  $\tilde{C}_{PT}$ must be real-valued while  $\tilde{C}_{CL}$ must be complex-valued;  (II) as compared with $\tilde{N}_{CL}$, $\tilde{N}_{PT}$ contains an additional term being the correlation function of the real and imaginary parts of the scattering potential.

\section{Application to Schell-model scatterers with linear and quadratic phases}

Let us assume that $C_{PT}$ of the scatterer is given by Eq. \eqref{CPTFTkernel}
where $a(\vec{r},\omega)$ and $\mu(\vec{r}_d,\omega)$ are selected as 
\begin{equation}\label{GSM}
\begin{split}
    a(\vec{r},\omega) &= I_0\exp\left( -\frac{r^2}{2a^2} \right) \exp(-i\vec{\alpha}\cdot\vec{r}), \\
    \mu(\vec{r}_d,\omega) &= \exp\left[ -\frac{(\vec{r}_1-\vec{r}_2)^2}{2d^2} \right]\exp[i\vec{\beta}\cdot(\vec{r}_1-\vec{r}_2)],
\end{split}
\end{equation}
where $I_0$ is a constant, $\vec{\alpha}$, $\vec{\beta}$ are real vectors, the former relating to the non-Hermiticity of the material's realizations and the latter characterizes non-Hermiticity of the material's correlation function. Further,
$a$ is related to the dimensions of the scatterer and $d$ is the correlation length. In the case of linear phases the effects of $\vec{\alpha}$ and $\vec{\beta}$ can be combined:

\begin{equation}
\vec{\gamma}=\vec{\alpha}+\vec{\beta}.    
\end{equation}
Note that if $\vec{\alpha}=-\vec{\beta}$ then $C_{PT}$ is symmetric and real valued and will result in axially symmetric scattered spectral density, even though the realizations of the material are $\mathcal{PT}$-symmetric. Such simple phase cancellation is not possible in cases when the phase functions are non-linear.  

Then using Eqs. \eqref{GSM} in Eq. \eqref{CPTFTkernel} and applying three-dimensional Fourier transform of $C_{PT}(\vec{r}_1, \vec{r}_2,\omega)$ yields
\begin{equation}\label{Clinear}
\begin{split}
    \tilde{C}_{PT}(-\vec{K}_1,\vec{K}_2,\omega) &= I_0^2 \frac{(2\pi)^{3/2}a^6 d^3 }{(2a^2 + d^2)^{3/2}} 
    \exp\left(\frac{ - a^2 \vec{K}_{1}\cdot\vec{K}_{2}}{2 + d^2/a^2} \right)
    \\
    &\times \exp\left[ \frac{-(a^2+d^2)(\vec{K}_{1}^2 + \vec{K}_{2}^2)/2}{2 + d^2/a^2}\right] \\
    &\times\exp\left[\frac{ - d^2\vec{\gamma}\cdot(\vec{K}_{1} + \vec{K}_{2} - \vec{\gamma})}{2 + d^2/a^2}\right].
\end{split}
\end{equation}
We confirm, based on the results of the previous section, that $\tilde{C}_{PT}$ in \eqref{Clinear} is real-valued. Also, in the limit $d \rightarrow \infty$, the scattering results pertinent to deterministic $\mathcal{PT}$-symmetric media can be deduced. Indeed, it is implied by Eq. \eqref{Clinear} that as $d \rightarrow \infty$ 
\begin{equation}
\begin{split}
    \tilde{C}_{PT}&(-\vec{K}_1,\vec{K}_2,d\rightarrow \infty,\omega) = I_0^2 (2\pi)^{3/2}a^6 \\
    &\times \exp\left\{ -\frac{a^2}{2} \left[ \vec{K}_1^2 + \vec{K}_2^2 -2\vec{\gamma}\cdot(\vec{K}_{1} + \vec{K}_{2} - \vec{\gamma}) \right] \right\},
\end{split}
\end{equation}
as expected. Expression \eqref{Clinear} can be also written in form evidently demonstrating its non-Hermitian character:
\begin{equation}
\begin{split}
    \tilde{C}_{PT}(-\vec{K}_1,\vec{K}_2,\omega) &= \tilde{C}_{PT}(-\vec{K}_1,\vec{K}_2,\vec{\gamma} = 0,\omega) \\
    &\times\exp\left[ \frac{ \vec{\gamma}\cdot(\vec{K}_{1} + \vec{K}_{2} - \vec{\gamma}) }{2/d^2 + 1/a^2} \right],
\end{split}
\end{equation}
where $\tilde{C}_{\vec{\gamma} = 0}(-\vec{K}_1,\vec{K}_2,\omega)$ is the correlation function for a Hermitian scatterer with $\vec{\gamma} = 0$, given explicitly by
\begin{equation}
\begin{split}
    \tilde{C}_{PT}&(-\vec{K}_1,\vec{K}_2,\vec{\gamma} = 0,\omega) = I_0^2 \frac{(2\pi)^{3/2}a^6 d^3 }{(2a^2 + d^2)^{3/2}} \\
    &\times\exp\left[ -\frac{(a^2+d^2)(\vec{K}_{1}^2 + \vec{K}_{2}^2)/2 - a^2 \vec{K}_{1}\cdot\vec{K}_{2}}{2 + d^2/a^2} \right].
\end{split}
\end{equation}
In particular, case $\vec{\alpha}\neq 0$ and $\vec{\beta}=0$ can be associated with the situation when the scatterer is deterministic and $\mathcal{PT}$-symmetric and is randomized by a real-valued correlation function. On the other hand, if $\vec{\beta}\neq 0$ but $\vec{\alpha}=0$ can be thought as a classic deterministic medium, but it still produces $\mathcal{PT}$-symmetry like effects. In principle, one can have a match  $\vec{\alpha}=-\vec{\beta}$ or $\vec{\gamma}=0$. The resulting field cannot be distinguished from a classic field, even though it would scatter from $\mathcal{PT}$-symmetric medium. This is a very special case of the discussion after Eq. \eqref{CPTFTkernel} for both linear phases. It appears impossible to arrange for such a match if at least one of the phases is not linear. 

The spectral density produced on scattering of the plane wave to the far-zone of the $\mathcal{PT}$-symmetric medium can be deduced from Eq. \eqref{Clinear} as
\begin{equation}\label{Ss}
\begin{split}
S^s_{PT}(r\hat{s},\omega) &= I_0^2\frac{S^i(\omega)}{r^2} \frac{(2\pi)^{3/2}a^6 d^3 }{(2a^2 + d^2)^{3/2}} \\
&\times\exp\left( -\frac{\gamma^2}{\sigma^2} \right)\exp\left(  \frac{-\vec{K}^2 + 2\vec{\gamma}\cdot\vec{K}}{\sigma^2}  \right), 
\end{split}
\end{equation}
where $\sigma^2 = 2/d^2 + 1/a^2$. Figures 2 and 3 give numerical examples of $S^s_{PT}$ in Eq. \eqref{Ss} for several values of $a\vec{\gamma} = (a\gamma_x,a\gamma_y,a\gamma_z)$ depending on polar angle $\phi$ and azimuthal angle $\theta$ of the spherical coordinate system, defined as:
\begin{equation}
s_x=\sin\theta\cos\phi, \quad  s_y=\sin\theta \sin\phi, \quad s_z=\cos\theta. 
\end{equation}

\begin{figure}[ht]
    \centering
    \includegraphics[width=0.45\textwidth]{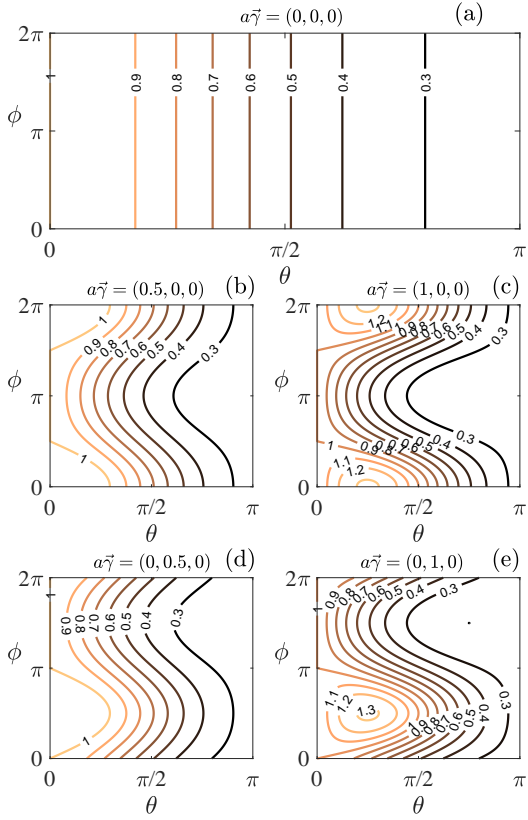}
    \caption{Contour plot of the position-dependent term in the spectral density \eqref{Ss} of the scattered radiation field for (a) $a \vec{\gamma} = (0,0,0)$, (b) $ a\vec{\gamma} = (0.5,0,0)$ and (c) $a\vec{\gamma} = (1,0,0)$, (d) $a\vec{\gamma} = (0,0.5,0)$ and (e) $a\vec{\gamma} = (0,1,0)$. Parameters used: $ka = 1$, $d/a = 1$ and incident direction $\hat{s}_0 = \hat{z}$. }
    \label{fig2}
\end{figure}

\begin{figure}[ht]
    \centering
    \includegraphics[width=0.326\textwidth]{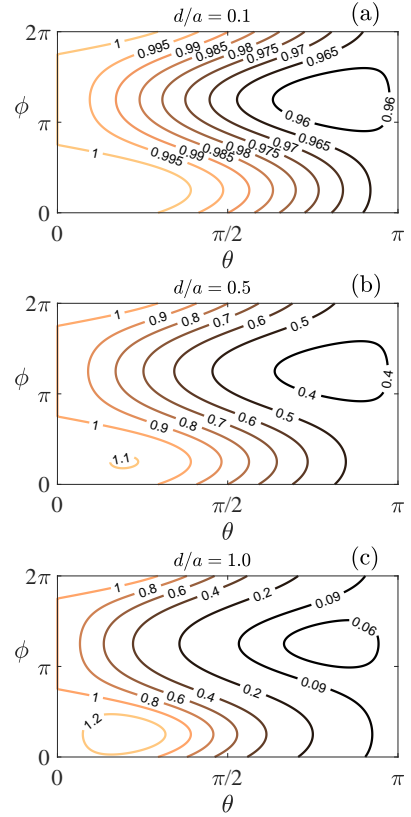}
    \caption{Contour plot of the position-dependent term in the spectral density \eqref{Ss} of the scattered radiation field for (a) $d/a = 0.1$, (b) $d/a = 0.5$ and (c) $d/a = 1$. Parameters used: $ka = 1$, $a\vec{\gamma} = (1,1,1)$ and incident direction $\hat{s}_0 = \hat{z}$.}
    \label{fig3}
\end{figure}

Consider now an example relating to classic medium. The $C_{CL}$ of the classic scatterer directly corresponding to Eq. \eqref{GSM} would have the form
\begin{equation}
\begin{split}
    &a(\vec{r},\omega) = I_0\exp\left( -\frac{r^2}{2a^2} \right) \exp[-i(\alpha_x |x|+\alpha_{y} |y|+\alpha_{z}|z|)], \\& \mu(\vec{r}_d,\omega) = \exp\left[ -\frac{r_d^2}{2d^2} \right]\exp[i(\beta_x |x_d|+\beta_y |y_d|+\beta_z |z_d|)].
\end{split}
\end{equation}
However, the Fourier transforms of $a$ and $\mu$ do not lead to simple analytic equations for $\tilde{C}_{CL}$. Instead we set 
\begin{equation}\label{aclas}
    a(\vec{r},\omega) = I_0\exp\left( -\frac{r^2}{2a^2} \right) \exp(-i\alpha r^2)
\end{equation}
while keeping $\mu_{CL}$ as in Eq. \eqref{GSM} with $\vec{\beta}=0$. Such classic medium with quadratic phase yields the complex-valued $\tilde{C}_{CL}$ of the form:
\begin{equation}\label{Cclassic}
\begin{split}
    \tilde{C}_{CL}(-\vec{K}_1,\vec{K}_2,\omega) &= I_0^2\left(\frac{2\pi}{c_0}\right)^{3/2}\left( \frac{c_1^2 + c_2^2}{c_3^2 + c_4^2}  \right)^{-3/4} \\
    &\times\exp\left[ \frac{-(a^2+d^2)(\vec{K}_1^2+\vec{K}_2^2)}{4 + 8\alpha^2 a^2 d^2 + 2d^2/a^2}\right] \\
    &\times \exp\left[\frac{ 2i\alpha a^2d^2(\vec{K}_2^2 - \vec{K}_1^2)  }{4 + 8\alpha^2 a^2 d^2 + 2d^2/a^2} \right] \\
    &\times \exp\left[\frac{2a^2\vec{K}_{1}\cdot\vec{K}_{2}}{4 + 8\alpha^2 a^2 d^2 + 2d^2/a^2} \right],
\end{split}
\end{equation}
where $c_0 = 4\alpha^2 a^4 d^2 + 2a^2 + d^2$, $c_1 = 1/d^{2}+1/a^{2}$, $c_2 = -2\alpha$, $c_3 = 2a^4 d^2 \alpha$ and $c_4 = \alpha^4 + a^2d^2$. Equation \eqref{Cclassic} clearly satisfies the condition $\tilde{C}_{CL}(-\vec{K}_2,\vec{K}_1,\omega) = \tilde{C}_{CL}^*(-\vec{K}_1,\vec{K}_2,\omega)$, as expected. Also, $\tilde{C}_{CL}$ is a complex-valued function but $\tilde{C}_{PT}(-\vec{K},\vec{K},\omega)$ is  real-valued. The spectral density produced on scattering from the classic medium, obtained from \eqref{Cclassic}, is given by
\begin{equation}
\begin{split}
    S_{CL}^s(r\hat{s},\omega) &=  \frac{I_0^2 S^i(\omega)}{r^2}\left(\frac{2\pi}{c_0}\right)^{3/2}\left( \frac{c_1^2 + c_2^2}{c_3^2 + c_4^2}  \right)^{-3/4} \\
    &\times \exp\left( \frac{-\vec{K}^2}{4\alpha^2 a^2 + 2/d^2 + 1/a^2} \right),
\end{split}
\end{equation}
and it depends only on the angle between $\hat{s}_0$ and $\hat{s}$. This is in contrast to the spectral density for the $\mathcal{PT}$-symmetric medium where it depends not only on $\vec{K}^2$ but also on $\vec{\gamma}\cdot\vec{K}$ [see Eq. \eqref{Ss}].

\begin{figure}[ht]
    \centering
    \includegraphics[width=0.32\textwidth]{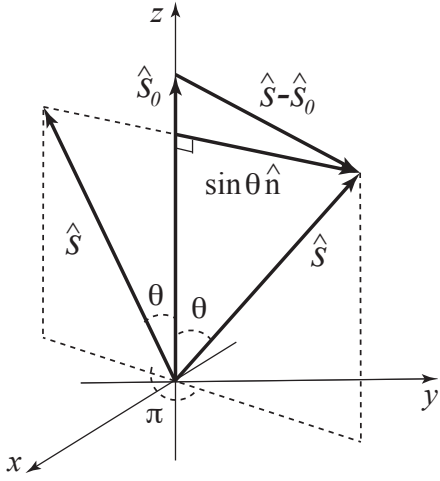}
    \caption{Geometry for the symmetrical scattered directions where $\hat{s}_0$ is the direction of the incident field and $\theta$ is the angle between $\hat{s}$ and $\hat{s}_0$. }
    \label{fig4}
\end{figure}

\begin{figure}[ht]
    \centering
    \includegraphics[width=0.45\textwidth]{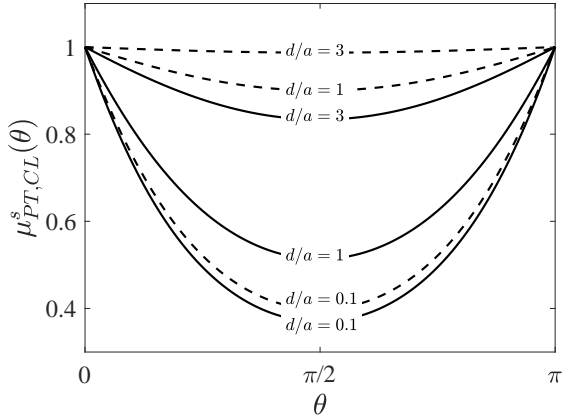}
    \caption{Spectral degree of coherence $\mu_{PT,CL}^s (\theta)$ for radiation scattered by $\mathcal{PT}$-symmetric (continuous lines) and classic (dashed lines) materials. Three values of $d/a$ are considered: 0.1, 1 and 3. Parameters: $ka = 1$ and $\alpha/k^2 = 2$.  } 
    \label{fig5}
\end{figure}

In closing this section, let us analyze the spectral degree of coherence $\mu^s(r\hat{s}_1,r\hat{s}_2,\omega)$ for the radiation field scattered by classic and $\mathcal{PT}$-symmetric media. Direct substitution of Eqs. \eqref{Clinear} and \eqref{Cclassic} into Eq. \eqref{mus} gives
\begin{equation}
\begin{split}
    \mu_{CL}^s(r\hat{s}_1,r\hat{s}_2,\omega) &= \exp\left[ \frac{2i\alpha a^2 d^2(K_2^2 - K_1^2)}{4 + 8\alpha^2 a^2 d^2 + 2d^2/a^2}\right] \\
    &\times\exp\left[\frac{-a^2(\vec{K}_1 - \vec{K}_2)^2}{4 + 8\alpha^2 a^2 d^2 + 2d^2/a^2} \right],
\end{split}
\end{equation}
and
\begin{equation}
    \mu^s_{PT}(\hat{s}_1,\hat{s}_2,\omega) = \exp\left[ -\frac{a^2(\vec{K}_1 - \vec{K}_2)^2}{2 + d^2/a^2} \right].
\end{equation}
It is usually the case where a pair of symmetrical directions are chosen to highlight the statistical properties of the scattered radiation, for example (see Fig. 4) 
\begin{equation}\label{schoice}
    \hat{s}_1 = \hat{s}, \quad \hat{s}_2 =
\hat{s}-2 \hat{n} \sin\theta,
\end{equation}

The symmetry suggested by these vectors depends only on the angle between $\hat{s}$ and $\hat{s}_0$ and is independent of the coordinate system. In the particular, basis-dependent, situation where $\hat{s}_0 = \hat{z}$, angle $\theta$ is the azimuthal angle in spherical polar coordinates. For this choice, 

\begin{equation}\label{musymm}
\begin{split}
\mu^s_{CL}(\theta,\omega) &=  \exp\left( \frac{ - a^2 k^2\sin^2\theta}{1 + d^2/2a^2 + 2\alpha^2 a^2 d^2} \right), \\
\mu^s_{PT}(\theta,\omega)  
    &= \exp\left( \frac{-a^2k^2\sin^2\theta}{1 + d^2/2a^2} \right).
\end{split}
\end{equation}
We thus found, for these particular symmetrical directions, that the spectral degree of coherence for the $\mathcal{PT}$-symmetric scatterers depends only on its geometrical properties while for the classic medium it depends on the phase $\alpha$ [see Eq. \eqref{aclas}]. Figure \ref{fig5} shows how $\mu_{CL,PT}^s(\theta,\omega)$, given by \eqref{musymm}, vary as a function of $\theta$ for several values of $d/a$.

\section{Summary} 

We have introduced a class of random, stationary media whose second-order spatial correlation functions of scattering potential obey the conditions of $\mathcal{PT}$-symmetry and have derived their major properties. In such media the balance between gain and loss centers can be achieved by requiring it from the individual realizations of the scattering potential (or, alternatively, the index of refraction). Also, individual potential's realizations can be passive (no gain or loss present) but the correlation function might possess a phase term leading to $\mathcal{PT}$-symmetric like effects. In order to illustrate the distinctive nature of the introduced media we have made a close comparison of their properties with those of ``classic'' media, i.e., geometrically symmetric media with  unbalanced gain and loss. Then we have applied the Bochner theorem of functional analysis for analytical modeling of the genuine $\mathcal{PT}$-symmetric correlation functions, and considered in detail a particular but very important subclass of such media with uniform (Schell-like) correlations.

Further we have developed a theory of light scattering from the $\mathcal{PT}$-symmetric media and applied it to an example of a plane wave interacting with Gaussian Schell-model media with linear phase functions. Such an example provides the required insight into the nature of light-medium interactions while offering unsurpassed simplicity. In particular, it illustrates the fact that both types of phase terms (of potential's realizations and of correlation function)   responsible for asymmetric intensity scattering can be present, the combination either enhancing the produced asymmetry or suppressing (annihilating) it.      



\end{document}